\documentclass{svproc}
\usepackage{url}
\usepackage{graphicx}
\usepackage[misc]{ifsym}
\usepackage{color}
\usepackage{amsmath}
\usepackage{amssymb}

\begin{document}
\mainmatter              
\title{Reconciling $B$-meson Anomalies, Neutrino Masses and Dark Matter}
\titlerunning{Reconciling $B$-meson anomalies}  
%
\author{ Girish Kumar\,\inst{1} (\Letter\,)  \and Chandan Hati\,\inst{2}\and
Jean Orloff\,\inst{2} \and Ana M. Teixeira\,\inst{2}}
\authorrunning{Girish Kumar et al.} 
%
\tocauthor{Chandan Hati, Girish Kumar, Jean Orloff, and Ana M. Teixeira}
\institute{Tata Institute of Fundamental Research, Homi Bhabha Rd., 400005 Mumbai, India,\\
\email{girishk@theory.tifr.res.in}
\and
Laboratoire de Physique de Clermont, CNRS/IN2P3 - UMR 6533,\\ 4 Avenue Blaise Pascal, F-63178 Aubi\`{e}re, France
}

\maketitle              

\begin{abstract}
We explore the connection of the leptoquark solution to the recently reported $B$-meson anomalies with a mechanism of neutrino mass generation and a viable dark matter candidate. We consider a model consisting of two scalar leptoquarks  and three generations of triplet fermions: neutrino masses are radiatively generated at the 3-loop level and, by imposing a discrete $Z_2$ symmetry, one can obtain a viable dark matter candidate. We discuss the constraints on the flavour structure of this model arising from numerous flavour observables. The rare decay $K\to \pi^+\nu\bar\nu$ and charged lepton flavour violating $\mu-e$ conversion in nuclei are found to provide the most stringent constraint on this class of models.
\end{abstract}
\section{Introduction}
Following the observation of the Higgs boson at the LHC, which was the last missing piece of the standard model (SM) to be discovered, a strong effort is being made to carry out tests of the SM, and unveil the presence of new physics (NP). The observation of neutrino oscillations and the cosmological evidence for dark matter in the Universe are among some of the most important reasons to believe that SM is a low-energy limit of a more fundamental theory, realised at some unknown high scale. Recently, the observed lepton flavour universality violation (LFUV) in $B$-decays has provided new hints of NP. In particular, measurements of the following ratios show significant deviations from the SM predictions
\begin{equation}
R_{K^{(\ast)}} \,= \,
\frac{{\rm BR}(B \to K^{(*)}\, \mu^+\,\mu^-)}{{\rm BR}(B \to K^{(*)}\, e^+\,e^-)}\,~ {\rm and}~
R^{\tau/\ell}_{D^{(*)}} \,= \,\frac{{\rm \Gamma}(B \to D^{(*)} \,\tau^- \,\bar\nu)}{{\rm \Gamma}(B \to  D^{(*)}\, {\ell^- }\,\bar\nu)}\,;~ \ell=e,\mu\,.
\end{equation}
The measured values of $R_{K^{(\ast)}}$, and the corresponding predictions in the SM in low dilepton invariant mass squared $q^2$ bins are as follows  \cite{Aaij:2014ora,Aaij:2017vbb,Bordone:2016gaq}
\begin{eqnarray}
& R_{K [1,6]}\, =\, 0.745\,\pm_{0.074}^{0.090}\,\pm\,
0.036,\, \quad
& R_{K}^{\rm{SM}}\, =\, 1.00, \pm\, 0.01 
 \nonumber\\
& R_{K^*[0.045, 1.1]}\, =\, 0.66 ^{+0.11}_{-0.07} \,\pm\,
 0.03\,, \quad  
&R_{K^*[0.045, 1.1]}^{\rm{SM}}\,  \sim\,  0.92\pm 0.02 \nonumber\\
& R_{K^* [1.1, 6]} \,= \,0.69 ^{+0.11}_{-0.07}\, \pm
0.05\,, \quad
& R_{K^* [1.1, 6]}^{\rm{SM}} \,\sim \, 1.00\pm 0.01\, ,
\end{eqnarray}
which respectively correspond to $2.6\sigma$, $2.4\sigma$ and $2.5\sigma$ deviations from the theoretical expectations for $R_{K [1,6]}$, $R_{K^*[0.045, 1.1]}$ and $R_{K^* [1.1, 6]}$. 
  On the other hand, experimental values for LFUV ratios $R^{\tau/\ell}_{D^{(\ast)}}$ are larger than the predictions in the SM. The current experimental world averages are $R_D^{\tau/\ell} = 0.407 \pm 0.039 \pm 0.024$ and $R_{D^\ast}^{\tau/\ell} = 0.306 \pm 0.013 \pm 0.007 $,  corresponding to a $2.3\sigma$ and $3.0\sigma$ excess over the respective SM values \cite{Amhis:2016xyh}. If combined together, the discrepancy is at the level of  $3.8\sigma$. Interestingly, measurements of $R_{D^{(\ast)}}^{e/\mu}$ (related to $e/\mu$ modes) do not show any sign of LFUV, and are consistent with the SM predictions.
   \begin{table}\label{Table1}
\caption{New fields (in addition to the usual SM fields) and their transformation under the SM gauge
  group and  discrete symmetry $Z_2$. All the SM fields are even under $Z_2$.}
\begin{center}
\begin{tabular}{cclr}
\hline
&\hspace*{5mm} Field	&\hspace*{5mm} {\small SU(3)$_C\times$SU(2)$_L\times$U(1)$_Y$ } & \hspace*{5mm}$Z_{2}$ \\
\hline\rule{0pt}{12pt}
Fermions &\hspace*{5mm} $\Sigma_R$ & \hspace*{15mm} $(\textbf{1}, \textbf{3},  0)$ & $-1$\\ 
\hline
\rule{0pt}{12pt}
Scalars	& \hspace*{5mm}$H$ & \hspace*{15mm} $(\textbf{1}, \textbf{2},1/2)$	&   $1$	\\ 
&\hspace*{5mm} $h_1$	& \hspace*{15mm} $(\bar{\textbf{3}}, \textbf{3}, {1/3})$ &   $1$\\ 
&\hspace*{5mm} $h_2$	&  \hspace*{16mm}$(\bar{\textbf{3}}, \textbf{3}, {1/3})$ &   $-1$ \\   
[2pt]
\hline
\end{tabular}
\end{center}
\end{table}
 Here, we discuss an extension of the SM by two scalar triplet leptoquarks $h_1$, $h_2$ and three generation of triplet fermions $\Sigma_R^i$. The corresponding quantum charges under the SM gauge group  {\small SU(3)$_C\times$SU(2)$_L\times$U(1)$_Y$} and an additional discrete symmetry $Z_2$ (which ensures the stability of the dark matter candidate) are listed in Table~\ref{Table1}.  As we proceed to discuss, this model has the potential to explain the neutral $B$-meson decay anomalies, neutrino oscillation data, and also provide a suitable dark matter candidate.  We also discuss possible contributions to several flavour observables such as neutral meson mixing, rare meson decays, and charged lepton flavour violating (cLFV) decays, and study the constraints arising from these processes on the model. For a detailed analysis we refer to the original work \cite{Hati:2018fzc}. 
 
 The relevant interactions in the Lagrangian are
\begin{eqnarray}\label{lag1}
\mathcal{L} \, &=& \, \mathcal{L}_{\rm SM} \,+\, 
y_{ij}\, \bar{Q}_{L}^{C\,i} \,i\tau^2 \,(\vec{\tau}. \vec{h}_{1})
\,L_{L}^{j}\,+\,
\tilde{y}_{ij}\,\overline{(\vec{\tau}. \vec{\Sigma})}^{C\,i,ab}_{R}
      [i\tau^2 \,(\vec{\tau}. \vec{h}_{2})\,\epsilon^{T}]^{ba}\,
      d_{R}^{j}\nonumber\\
  &&~~~~~~~~~~~~~~~~~~    -\frac{1}{2}\,\overline{\Sigma^{C}}^{i} \,M_{ij}^{\Sigma}\, \Sigma^{j} - \,
V^{H,h}_\text{scalar}
\, +\,\text{H.c.} \,,
\end{eqnarray}
where $Q_L $ is the left-handed quark doublet, $L_L$ is the left-handed lepton doublet and $d$ is the right-handed singlet down-type quark. Here $i,j$ are labels for   generations,  $a,b$ denote SU(2) indices and  $\tau^c$ ($c=1,2,3$) are the Pauli matrices. In writing the above Lagrangian, we have forbidden the di-quark interaction of $h_1$ to be consistent with proton decay bounds. The first term in the second line is the mass term for fermion triplets $\Sigma_R^i$, and $V^{H,h}_\text{scalar}$ is the scalar potential. The full expression of the scalar potential can be found in Ref.~\cite{Hati:2018fzc}. In particular,  it contains a quadratic term ($\lambda_h/4$)\,$\text{Tr}(h_1^{\dagger} \,h_2 \,h_1^{\dagger} \,h_2)$ which is instrumental for generating neutrino masses in this model. 
\section{Radiative Neutrino Masses and Dark Matter}
In this model, the $Z_2$ symmetry forbids any tree level realisation of the conventional seesaw mechanism. However, neutrino masses are generated radiatively and the leading contribution arises at three loop level. 
After calculating all the relevant loop diagrams, one obtains 
\begin{equation}{\label{eq:numass}}
(m_{\nu})_{\alpha\beta}\, =\, -30\, 
\frac{\lambda_{h}}{\left(4\pi^{2}\right)^{3}\, m_{h_{2}}}\, y^{T}_{\alpha i}\, 
m_{D_{i}} \, \tilde{y}^{T}_{ij} \, 
G\left(\frac{m_{\Sigma_{j}}^{2}}{m_{h_{2}}^{2}},
\frac{m_{h_{1}}^{2}}{m_{h_{2}}^{2}}\right)\, 
\tilde{y}_{jk}\, m_{D_{k}}\, y_{k \beta}   ,
\end{equation}
where $y$ and $\tilde{y}$ are Yukawa couplings which have been defined in eq.~\eqref{lag1}, $m_D$ denotes the diagonal down-quark mass matrix and $G(a,b)$ is a loop function given in Ref.~\cite{Hati:2018fzc}. The charged leptons are assumed to be in the physical basis and the Pontecorvo-Maki-Nakagawa-Sakata (PMNS) matrix, $U_{ij}$, parametrises the misalignment between flavour and mass eigenstates of the three SM neutrinos. The diagonal neutrino mass matrix is obtained using the relation $m_{\nu}^{\text{diag}}=\,U^{T}_{i\alpha}\, (m_{\nu})_{\alpha \beta} \,U_{\beta i}$. 

A numerical analysis of neutrino oscillation data can be  performed using a modified Casas-Ibarra parametrisation \cite{Hati:2018fzc} to obtain one of the Yukawa couplings in terms of other Yukawa coupling as
\begin{align}\label{eq:CI_param}
\tilde{y} \,= \, [F(\lambda_h,m_\Sigma,m_{h_{1,2}})]^{-1/2}\; \mathcal {R}\; \sqrt{m^\text{diag}_\nu} 
\;U^\dag y^{-1} m_d^{-1}\,.
\end{align}
Here $\mathcal {R}$ is an arbitrary complex orthogonal matrix and the function $F(\lambda_h,m_\Sigma,m_{h_{1,2}})$ can be found in Ref.~\cite{Hati:2018fzc}. \newline

Since all the SM fields are $Z_2$-even, the lightest $Z_2$-odd particle is stable in this model. Noting that the electroweak radiative corrections render the charged component slightly heavier than the neutral ones, the neutral component ($\Sigma^0$) of the lightest lepton triplet ($\Sigma_R^1$) emerges as a dark matter candidate. The relic density, predominantly governed by the gauge interactions, can be obtained as 
\begin{equation}\label{eq:relic}
\Omega \,h^{2}\,\simeq\,
\frac{1.07 \times 10^9 \,x_f }{g_{\ast}^{1/2}\; M_{\text{Pl}}({\text{GeV}})\; \langle \sigma_{\text{eff}}|\bar v|\rangle},
\end{equation}
where $x_{f}\equiv m_{\Sigma}/T_f$ is the freeze-out temperature and $\langle \sigma_{\text{eff}}|\bar v|\rangle$ is obtained by including the gauge coannihilation of the charged and neutral components of $\Sigma_R^1$ as described in \cite{Hati:2018fzc}. The numerical analysis of the relic abundance yields a rough benchmark limit for the  $\Sigma_R^1$ mass given by   $2.425\;{\text{TeV}}< m_{\Sigma}<2.465\;{\text{TeV}}$  to obtain the correct $\Omega\, h^{2}= 0.1186 \,\pm \,0.0020$. This can in turn be used as a first constraint on the parameter space of the model.
\section{$B$-meson decay anomalies}
The $B\to K^{(\ast)} \ell^+\ell^-$ decays are described by the low-energy effective Hamiltonian~\cite{Hiller:2017bzc}
\begin{equation}\label{eq:effH_RK}
\mathcal{H}_{\rm eff} 
\,= \,- \frac{4 G_F}{\sqrt{2}}\,\frac{\alpha_{em}}{4\pi}\,V_{tb}V_{ts}^\ast\left[C_9\,(\bar s_L\gamma^\mu b_L)(\bar\ell\gamma_\mu\ell) + C_{10}\,(\bar s_L\gamma^\mu b_L)(\bar\ell\gamma_\mu\gamma_5\ell) \right]+
\text{H.c.}\,,
\end{equation}
where only effective operators relevant for addressing LFUV in B-decays have been kept. New contributions to Wilson coefficients (WC) $C_9$ and $C_{10}$ are given~by
\begin{equation}\label{eq:bsll}
C_9^{\ell\ell^\prime} \,= \,-C_{10}^{\ell\ell^\prime}\, =\, 
\frac{\pi \,v^2}{\alpha_e \,V_{tb}\,V_{ts}^\ast} \,
\frac{y_{b\ell^\prime}\, y_{s\ell}^\ast}{m_{h_1}^2}\,.	
\end{equation}
where $v$ is the Higgs vacuum expectation value. Model-independent studies to explain LFUV in $b\to s \ell^+\ell^-$ decays advocate NP in left-handed currents, modifying $C_9$ and $C_{10}$ \cite{Hiller:2017bzc}. Adapting the NP solutions obtained in Ref.~\cite{Hiller:2017bzc} to our case, we obtain  $-0.7 \, \lesssim  \,
\text{Re}[C_{9, {\rm NP}}^{\mu \mu}\,- \,C_{9, {\rm NP}}^{e e}]
 \,\lesssim \, -0.4$, which translates to the following constraint (at $1\sigma$)
\begin{equation}\label{eq:mu.e.coupling.fit}
0.64 \times 10^{-3}\,\lesssim\, 
\frac{\text{Re}[y_{b\mu}\,y_{s\mu}^\ast \,-\, y_{be}\,y_{se}^\ast]}
{(m_{h_1}/1\text{TeV})^2}\, \lesssim \,1.12 \times 10^{-3}\,.
\end{equation}
In Fig.~\ref{fig:final} (left plot) we show the parameter space in the plane of the Yukawa couplings  $y_{b\mu}\,y_{s\mu}^\ast$ and  $y_{be}\,y_{se}^\ast$ satisfying the above condition to explain the $R_K$ and $R_{K^\ast}$ anomalies.  
We note that in this model the $h_1$  leptoquark also contributes to $b\to c \ell^-\bar\nu$ at tree level. After integrating out the $h_1$ leptoquark, we obtain the following effective NP Hamiltonian
 \begin{align}\label{eq:effHRD}
\mathcal{H}_\text{eff}(b\to c\ell\bar \nu_i) =- \frac{v^2\,(yU)_{3i}\,(Vy^\ast)_{2\ell}\,}{4 \,V_{cb}\,m_{h_1}^2}\,
\left(\bar c_L\gamma^\mu b_L\right)
\left(\bar \ell_L \gamma_\mu\nu_{Li}\right)+ \text{H.c.}\;.
\end{align} 
Once various flavour constraints are taken into account, one finds SM-like $R_{D}$ and $R_{D^\ast}$ in this model. The LFUV ratios $R_{D^{(\ast)}}^{e/\mu}$, reflecting  $\mu/e$ universality, are also SM-like, and  consistent with current data.
 \begin{figure}[b!]
\begin{center}
\includegraphics[width=0.758\textwidth]{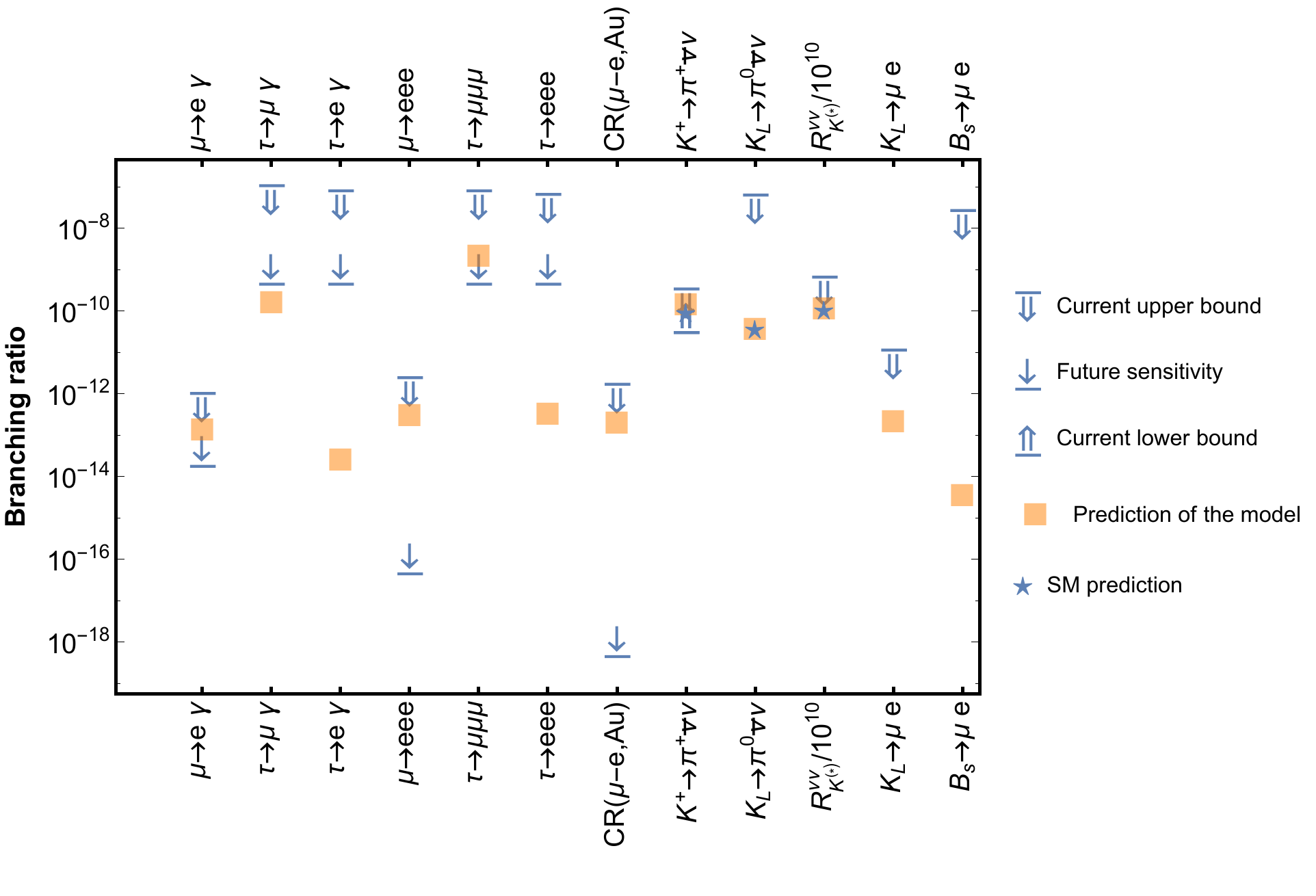}
\caption{Predictions of various processes for the flavour texture of the $h_1$ leptoquark,  as discussed in the text. The relevant information on the corresponding SM prediction and experimental data is also shown. For details see \cite{Hati:2018fzc}.}
\label{fig:summary}
\end{center}
\end{figure}
\section{Constraints from flavour data}
 We note that the simplest explanation of $b\to s \mu^+\mu^-$ anomalies requires nonzero values of the muon couplings $y_{b\mu}y_{s\mu}^\ast$ complying with eq.~\eqref{eq:mu.e.coupling.fit}. However, as evident from the relation in eq.~\eqref{eq:CI_param}, accommodating neutrino oscillation data implies a nontrivial flavour structure of the leptoquark $h_1$ Yukawa couplings, thereby inducing new contributions to a plethora of flavour processes. One can consider generic parametrisations of $y$ in terms of powers of
a small parameter $\epsilon$, with each
entry weighed by an $\mathcal{O}(1)$ real coefficient $a_{ij}$:
$y_{ij} \,=\, a_{ij} \phantom{|}_\odot \,\epsilon^{n_{ij}}\,,$ with $\odot$ denoting that there is no summation implied over $i,j$. After incorporating constraints from all the flavour violating processes, one can identify the allowed flavour textures which simultaneously explain $R_{K^{(\ast)}}$ while being consistent with all the constraints from flavour violating processes.  Among the  several textures analysed in Ref.~{\cite{Hati:2018fzc}}, one example of an allowed texture for the Yukawa couplings is given~by
{\small  \begin{align}
 y \,\simeq \,\left( 
\begin{array}{ccc}
\epsilon^{4} & \epsilon^{5} & \epsilon^{2} \\ 
\epsilon^{3} & \epsilon^{3} & \epsilon^{4} \\ 
\epsilon^{4} & \epsilon & \epsilon
\end{array}
\right)\,, \label{eq:gentexture}
\end{align}}
with $\epsilon \sim 0.215$ (for a leptoquark mass $m_{h_1} = 1.5$ TeV).  The value of the parameter $\epsilon$ is obtained from the $R_{K^{{(\ast)}}}$ anomaly constraint given in eq.~\eqref{eq:mu.e.coupling.fit}, by setting the product $y_{b\mu}y_{s\mu}^\ast$ $\sim \epsilon^4$ (with $\epsilon \sim \mathcal{O}(1)$). In Fig.~\ref{fig:summary} we show the predictions for some of the most important flavour violating processes for the texture given in eq.~\eqref{eq:gentexture}. A complete analysis suggests that $K^+\to \pi^+\nu\bar\nu$ and $\mu-e$ conversion in nuclei provide the most stringent constraints on the textures. 

\begin{figure}[t]
\begin{center}
\includegraphics[width=0.334\textwidth]{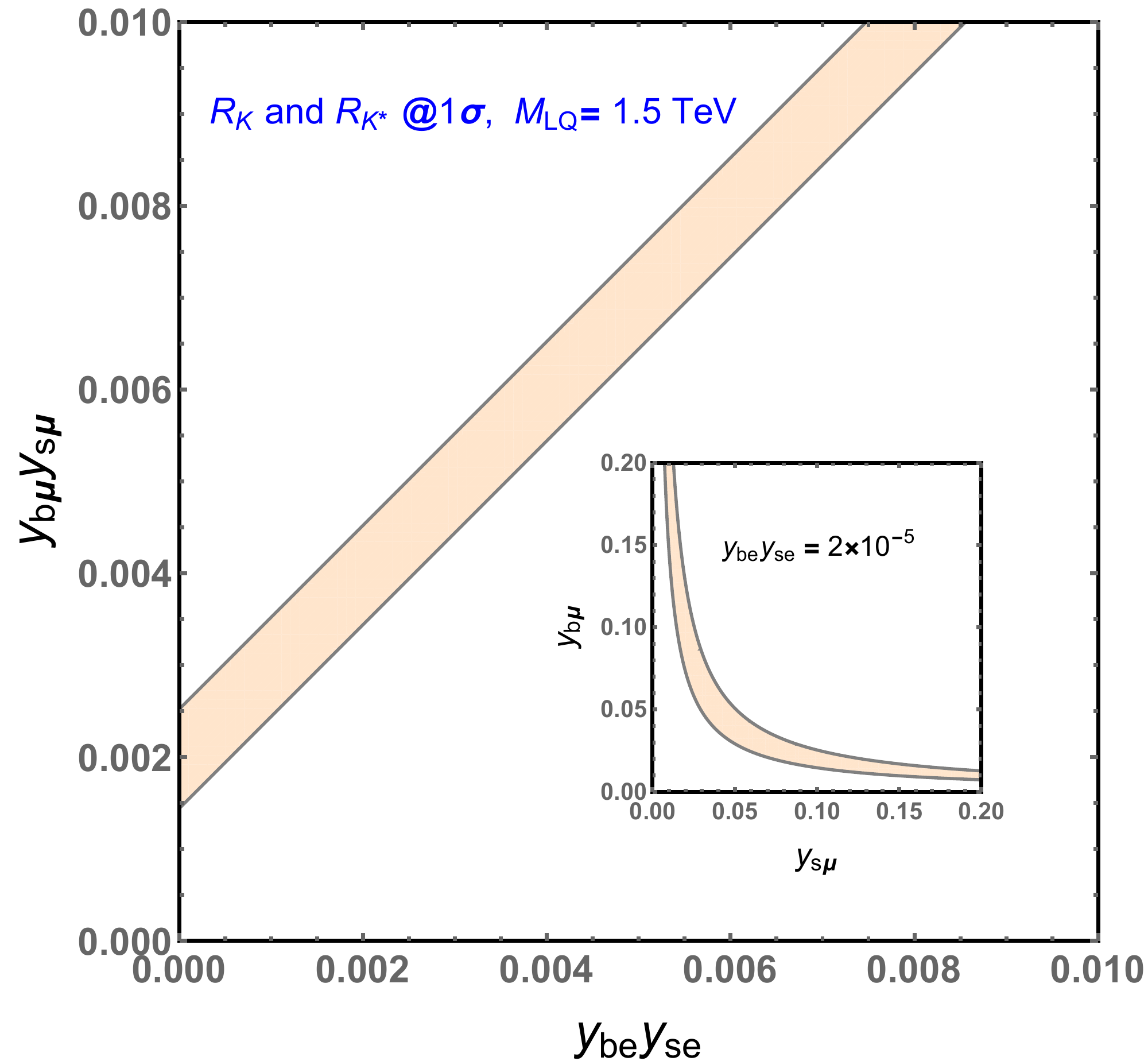}
\hspace{4ex}
\includegraphics[width=0.488\textwidth]{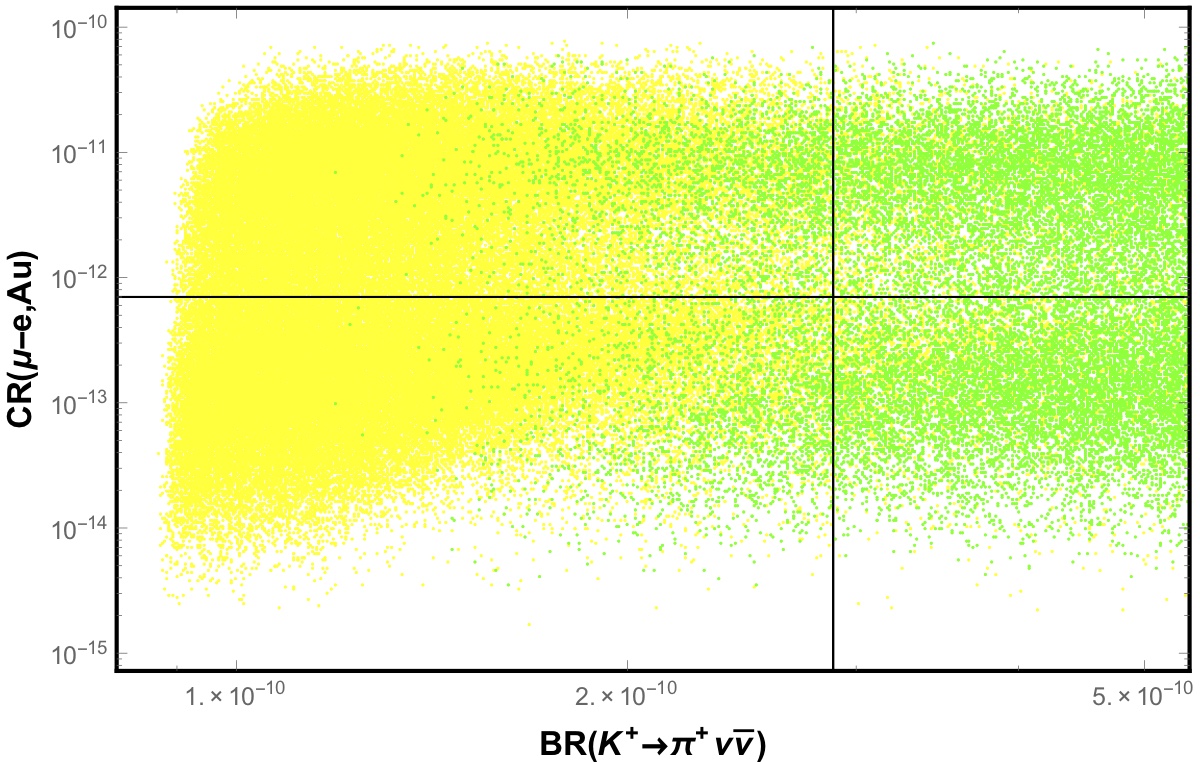}
\hspace{1ex}
\caption{The left plot shows the parameter space explaining $R_K$ and $R_{K^\ast}$ anomalies. In the right plot we show the allowed solutions from the neutrino oscillation data in the plane of BR($K^+\to \pi^+\nu\bar\nu$) and $\mu-e$ conversion rate for gold nuclei (for textures similar to eq.~\eqref{eq:gentexture}). The yellow(green) points correspond to non-perturbative (perturbative) Yukawa couplings $\tilde{y}$.}
\label{fig:final}
\end{center}
\end{figure}

In what concerns neutrino oscillation data, the best-fit values from the global oscillation analysis of~\cite{deSalas:2017kay} are used. Taking a normal ordering with the lightest neutrino mass in the range $[10^{-8}~\text{eV}, 0.001~\text{eV}]$, using the benchmark values $m_{\Sigma}$ = 2.45, 3.5 and 4.5~TeV, $m_{h_2} =$ 2.7~TeV, randomly sampling the three complex angles in $\mathcal{R}$ from the intervals: $[0,2\pi]$ for the phases, and $[-4\pi,4\pi]$ for the angles, one can obtain the couplings $\tilde y$ for different $y$ (for perturbative regimes $\lambda_h\lesssim 4\pi$ ). Finally, each entry of the couplings must comply with perturbativity requirements,  $|{y} (\tilde y)| \lesssim 4\pi$ and the various constraints from flavour violating processes must be satisfied. In Fig.~\ref{fig:final} (right plot) we show the results of the scan, for the texture given in eq. (\ref{eq:gentexture}) in the plane of neutrinoless conversion in nuclei and the 
$K^+ \to \pi^+ \nu \bar \nu$ decay; the colour code distinguishes between perturbative and
non-perturbative entries of the $y$ and $\tilde y$ couplings.
\section{Summary}
We discussed a model containing two leptoquarks $h_1$ and $h_2$ and  three Majorana triplets in addition to the SM fields. Neutrino masses are radiatively generated at the 3-loop level, and the neutral component of the lightest triplet is a viable dark matter candidate. The model allows to accommodate the observed anomalies in $B\to K^{(\ast)}\ell^+\ell^-$ decays, $R_K$ and $R_{K^\ast}$.  We discuss contributions to low energy flavour observables and cLFV processes and identify  processes which provide the tightest constraints on this class of models: these turn out to be $\mu-e$ conversion in nuclei and $K^+\to \pi^+\nu\bar\nu$ decays.  
%
%
%

%
\end{document}